\documentclass[aip,reprint]{revtex4-1}
\usepackage{graphicx} 
\usepackage[utf8]{inputenc}  
\usepackage[english]{babel}  

\usepackage{siunitx}
\begin{document}


\title{A laser-ARPES study of LaNiO$_3$ thin films grown by sputter deposition} 


\author{Edoardo Cappelli}
\email[Corresponding author: ]{edoardo.cappelli@unige.ch}
\thanks{This article has been submitted to APL Materials. After it is published, it will be found at \url{https://aip.scitation.org/journal/apm}.}
\affiliation{Department of Quantum Matter Physics, University of Geneva, 24 Quai Ernest-Ansermet, 1211 Geneva 4, Switzerland}
\author{Willem Tromp}
\affiliation{Department of Quantum Matter Physics, University of Geneva, 24 Quai Ernest-Ansermet, 1211 Geneva 4, Switzerland}
\affiliation{Leiden Institute of Physics, Leiden University, Niels Bohrweg 2, 2333 CA Leiden, The Netherlands}
\author{Siobhan McKeown Walker}
\affiliation{Department of Quantum Matter Physics, University of Geneva, 24 Quai Ernest-Ansermet, 1211 Geneva 4, Switzerland}
\author{Anna Tamai}
\affiliation{Department of Quantum Matter Physics, University of Geneva, 24 Quai Ernest-Ansermet, 1211 Geneva 4, Switzerland}
\author{Marta Gibert}
\affiliation{Department of Quantum Matter Physics, University of Geneva, 24 Quai Ernest-Ansermet, 1211 Geneva 4, Switzerland}
\affiliation{Department of Physics, University of Zürich, Winterthurerstrasse 190, 8057 Zürich, Switzerland}
\author{Felix Baumberger}
\affiliation{Department of Quantum Matter Physics, University of Geneva, 24 Quai Ernest-Ansermet, 1211 Geneva 4, Switzerland}
\affiliation{Swiss Light Source, Paul Scherrer Institut, CH-5232 Villigen PSI, Switzerland}
\author{Flavio Y. Bruno}
\affiliation{Department of Quantum Matter Physics, University of Geneva, 24 Quai Ernest-Ansermet, 1211 Geneva 4, Switzerland}
\affiliation{GFMC, Departamento de Fisica de Materiales, Universidad Complutense de Madrid, 28040 Madrid, Spain}


\date{\today}

\begin{abstract}
Thin films of the correlated transition-metal oxide LaNiO$_3$ undergo a metal-insulator transition when their thickness is reduced to a few unit cells. Here, we use angle-resolved photoemission spectroscopy to study the evolution of the electronic structure across this transition in a series of epitaxial LaNiO$_3$ films of thicknesses ranging from 19 to 2 u.c. grown \textit{in situ} by RF magnetron sputtering. Our data show a strong reduction of the electronic mean free path as the thickness is reduced below 5 u.c. This prevents the system from becoming electronically two-dimensional, as confirmed by the largely unchanged Fermi surface seen in our experiments. In the insulating state we observe a strong suppression of the coherent quasiparticle peak but no clear gap. These features resemble previous observations of the insulating state of NdNiO$_3$.
\end{abstract}

\maketitle 

The members of the rare-earth nickelate series $R$NiO$_3$ ($R$ = La, Pr, Nd...) undergo a first-order metal-insulator transition (MIT) from a high-temperature correlated metallic phase to a paramagnetic insulating phase followed by a transition to an antiferromagnetic ground state. The critical temperature of both transitions varies with the rare-earth element and correlates with the increasing distortion of the Ni$-$O$-$Ni bond angle with decreasing rare-earth ionic radius
~\cite{medarde1997,catalano2018}. The MIT is accompanied by a breathing distortion in which two inequivalent Ni sites with different Ni$-$O bond lengths are arranged in a three-dimensional (3D) checkerboard pattern. The resulting bond disproportionation leads to electron localization on one of the sublattices and can thus be seen as a site-selective Mott transition, which is strongly coupled to the lattice, while magnetic order seems to play a secondary role~\cite{park2012,peil2019}.

LaNiO$_3$ (LNO) represents an exception in the $R$NiO$_3$ family in that it remains metallic all the way down to the lowest temperatures. The metallic phase is on the brink of a MIT, and its low-temperature magnetic state is still under debate~\cite{sanchez1993, guo2018}.
Experiments on thin films have shown that it is possible to modulate the transport properties by electric-field gating~\cite{scherwitzl2009} and by tuning the strain on the lattice~\cite{son2010, chakhalian2011}. Moreover, a transition to an insulating state at all temperatures has been achieved by inducing oxygen vacancies~\cite{sanchez1996} or by reducing the thickness of the material down to a  few unit cells (u.c.)~\cite{scherwitzl2011, sakai2013, king2014, golalikhani2018}.
This latter thickness-dependent transition was originally ascribed to strong localization, based on evidence from transport experiments for a mean free path below the Ioffe-Regel limit $k_F \lambda = 1$, where $k_F$ is the Fermi wave vector and $\lambda$ the mean free path~\cite{scherwitzl2011}. It was later pointed out that structural distortions at the surface of LNO thin films, which have been observed both in scanning transmission electron microscopy (STEM) and X-ray diffraction experiments, could also play a role. In fact, both the top \mbox{LNO-vacuum} and the bottom \mbox{LNO-substrate} interfaces of thin films have been shown to be characterized by different conductivity and structural properties with respect to the interior~\cite{chen2013, kumah2014, fowlie2017, ruf2017}. Most strikingly, Kumah \textit{et al.}~\cite{kumah2014} were able to turn an insulating film metallic by adding a LaAlO$_3$ (LAO) capping layer, which suppressed the surface relaxation. This experiment demonstrated the key role of such relaxations and evidenced the fact that confined LNO layers in superlattices behave differently than the films with a free surface that we study here~\cite{kumah2014, golalikhani2018}. There is still no consensus regarding the nature of the insulating state in thin films, and the question of whether it is possible to stabilize a metallic 1-u.c.-thick LNO film remains open.

\begin{figure*}[ht]
\includegraphics[width=\textwidth]{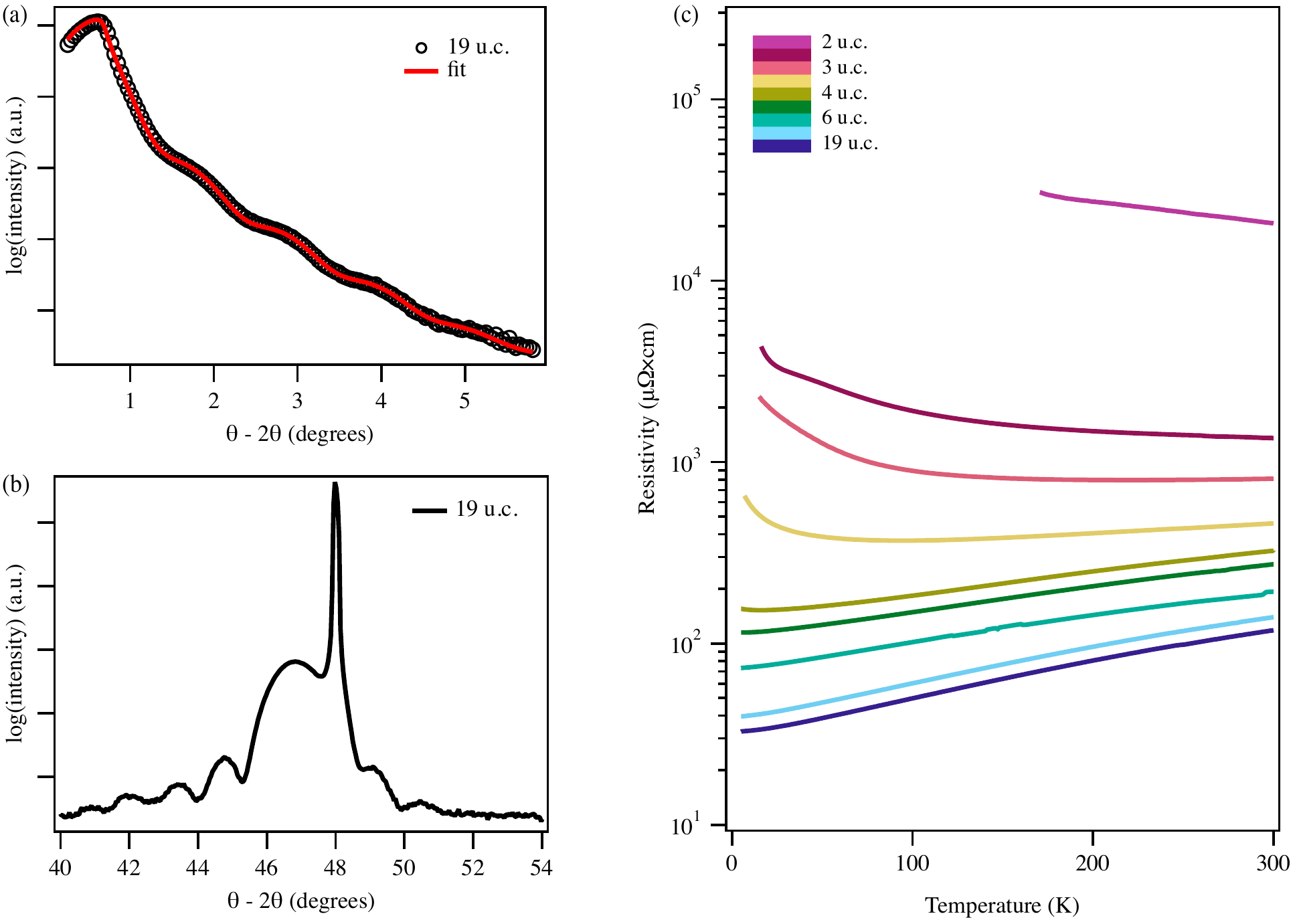}
\caption{\label{fig:x-ray+transport}(a,b) X-ray reflectivity and diffraction scans for a 19-u.c. LNO film. The fit indicated in red was obtained using the GenX software package~\cite{bjork2007}. (c) Resistivity measurements as a function of temperature for all LNO films. The thinnest LNO film could only be measured down to $\approx 150$~K before becoming too resistive for our 4-point setup.}
\end{figure*}

Over the past decade, several angle-resolved photoemission (ARPES) studies have been conducted on LNO epitaxial films grown by pulsed-laser deposition (PLD) and molecular-beam epitaxy (MBE). The Fermi surface of the correlated metallic phase has been found to be in good qualitative agreement with density functional theory (DFT) predictions for thick films, yet the reports are conflicting for thinner films~\cite{eguchi2009, king2014, yoo2016}. King \textit{et al.}~\cite{king2014} reported a 3D electronic structure as well as a MIT characterized by a loss of coherent quasiparticles at $E_F$ and a thickness-independent electronic mean free path, while Yoo \textit{et al.}~\cite{yoo2016} observed a modified Fermi surface for films thinner than 4 u.c., with signatures of a quasi-1D electronic structure. The former study attributed the onset of the insulating state to a fluctuating spin order enhanced by dimensionality, while the latter argued for the importance of 1D-Fermi-surface nesting in determining the MIT; a clear answer remains to be found. Some features of the spectral function of insulating LNO thin films are shared by thick insulating NdNiO$_3$ (NNO) films. Indeed, Dhaka \textit{et al.}~\cite{dhaka2015} reported a drastic reduction of spectral weight at $E_F$ in NNO at temperatures just below the metal-insulator transition, so the possibility of a common origin for the insulating states in LNO and NNO thin films is open. 

In this work, we report high-resolution laser-based ARPES measurements on a series of high-quality LNO films grown by sputter deposition. 
We show that the transition to increasingly insulating behavior with reduced thickness proceeds via a strong reduction in the quasiparticle mean free path and coherent weight, rather than the opening of a gap or a transition to electronically 2D behavior. 
We find that the spectral properties of the thinnest insulating samples are reminiscent of the low-temperature insulating state of NNO, suggesting a key role of structural relaxation at the surface. 

A series of LNO films with thicknesses ranging from 2 to 19 u.c. (\textit{i.e.} 7.5 to 71.8~\AA) were grown on LAO~(001) substrates by off-axis RF magnetron sputtering at a temperature~\mbox{$ T\approx 500^{\circ}$C} and pressure $P = 0.18$~mbar in a mixed 7:2 \mbox{Ar:O$_2$} atmosphere. In order to limit the formation of oxygen vacancies, the films were left to cool down in a pure oxygen atmosphere, maintaining the growth pressure.

\begin{figure*}[ht]
\includegraphics[width=\textwidth]{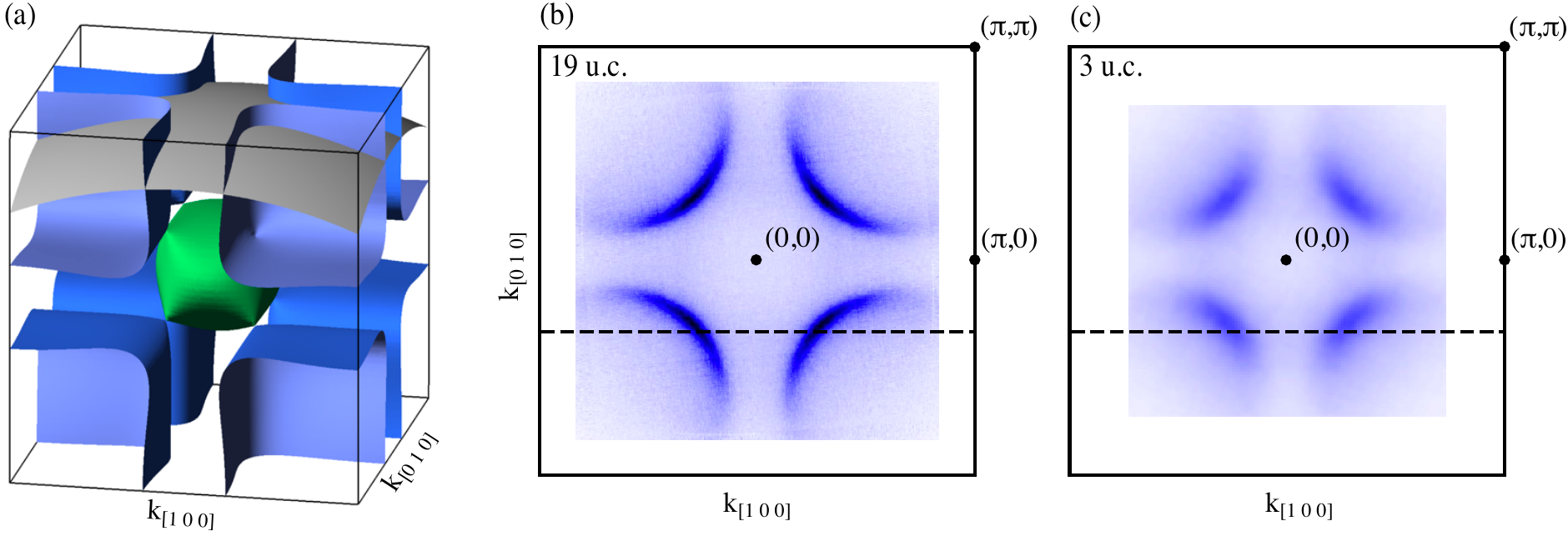}
\caption{\label{fig:fermi_surface}(a) Tight-binding calculation of the three-dimensional Fermi surface of bulk LNO, where the ratio between nearest- and next-nearest-neighbor hopping parameters has been taken from Ref.~\cite{lee2011}. The gray spherical contour indicates the momenta of free-electron final states for 11-eV photons and an inner potential of 15~eV, as determined in Ref.~\cite{bruno2017}. The hole pockets at the Brillouin-zone corners are shown in blue, and the electron pocket centered at $\Gamma$ is shown in green. (b,c) Experimental Fermi surfaces for films of 19- and 3-u.c. thickness (symmetrized). The vanishing intensity towards the edges of the Brillouin zone is due to matrix-element effects. The dashed line identifies the $k$-space cut along which the data in Figure~\ref{fig:arpes} were acquired.}
\end{figure*}

Immediately after growth, each film was transferred \textit{in vacuo} to the laser-ARPES setup so as to reduce surface contamination, which can affect the quality of the photoemission spectra. Photoemission experiments were performed with an electron spectrometer from MB Scientific and a narrow-bandwidth 11-eV pulsed laser from Lumeras, operated at a repetition rate of 50~MHz with 30~ps pulses of the fundamental infrared wavelength.  The energy resolution was \mbox{$\approx 7$~meV}, which was verified by fitting the Fermi edge of a polycrystalline gold sample, while the angular resolution was~\mbox{$\approx 0.003$~\AA$^{-1}$}. These values are well below the line widths observed in our experiments. 
The samples were mounted on a cryogenic 6-axes goniometer~\cite{hoesch2017} and measured at a temperature of \mbox{$\approx 5$~K} with the exception of the 2-u.c. film, which was measured at~\mbox{$\approx 50$~K} in order to limit sample charging due to its insulating nature.

The structural quality of the samples was characterized by x-ray reflectivity (XRR) and diffraction (XRD) \mbox{$\theta-2\theta$} scans using a Philips X'Pert 1 diffractometer equipped with a \mbox{Cu K$\alpha_1$} source ($\lambda = 1.5402$~\AA). Figures \ref{fig:x-ray+transport}(a) and (b) show XRR and XRD plots respectively for a 19-u.c.-thick sample. The reflectivity data was fitted using GenX in order to obtain the film thickness used to calibrate the growth rate and the surface roughness, which is below 1 u.c. for all films~\cite{bjork2007}. The XRD measurements show a sharp Bragg peak, corresponding to the LAO (002) planes, as well as a broader peak with side oscillations from the LNO thin film. The thickness obtained from these oscillations is consistent with the value extracted from the XRR fit. All in all, the quality of the XRD data and films is consistent with previous studies~\cite{fowlie2017}.

Platinum electrical contacts were deposited at the four corners of the samples. The sheet resistance was measured over a temperature range of 5 to 300~K in a Quantum Design Physical Property Measurement System using the standard van der Pauw configuration~\cite{vdpauw1958}. Figure \ref{fig:x-ray+transport}(c) shows resistivity curves as a function of temperature for all the LNO films. The residual resistivity ratios (RRR$\,\equiv\, \rho (300\textrm{K})/ \rho (5\textrm{K})$) for the metallic films (4$-$19 u.c.) vary between 3.5 and 2 and are comparable to the values measured for films grown by sputter deposition~\cite{scherwitzl2009, scherwitzl2011, fowlie2017} and pulsed-laser deposition (PLD)~\cite{horiba2007, golalikhani2018}, but 2-3 times lower than the RRR of the MBE-grown films measured in Ref.~\cite{king2014}. We can observe 3 different transport regimes: films thicker than 5 u.c. are metallic ($d\rho/dT>0$) all the way from room temperature down to the base temperature of \mbox{$\approx 5$K}. The resistivity of thinner films presents an upturn at low temperatures that has been observed before~\cite{scherwitzl2011, king2014, wang2019} and was attributed to weak localization effects~\cite{son2010, scherwitzl2011}. Finally, films thinner than 3 u.c. are insulating up to room temperature. 
These findings are fully consistent with previous reports for LNO films grown on LAO~\cite{king2014, fowlie2017, golalikhani2018}.

Figure \ref{fig:fermi_surface}(a) shows a tight-binding calculation of the three-dimensional Fermi surface of bulk LNO, which was performed using the parameters introduced by Lee \textit{et al.}~\cite{lee2011} and refined to fit ARPES measurements by Bruno \textit{et al.}~\cite{bruno2017}. The Fermi surface originates mainly from Ni $e_g$ orbitals and is composed of a nearly spherical electron pocket centred at $\Gamma$ and hole pockets at the Brillouin-zone corners. 
Figure \ref{fig:fermi_surface}(b) displays the photoemission intensity at the Fermi level obtained for the 19-u.c.-thick sample. 
We find well-defined nearly circular Fermi-surface contours centered at the Brillouin-zone corners. These contours have an arc-like appearance with a weight that rapidly decays away from the zone diagonal, reminiscent of underdoped cuprates~\cite{shen2005}. However, a careful inspection of the data does not show any signs of a momentum-dependent quasiparticle linewidth or of the opening of a pseudogap. 
We thus attribute the weight distribution to momentum-dependent matrix elements.

\begin{figure*}[ht]
\includegraphics[width=\textwidth]{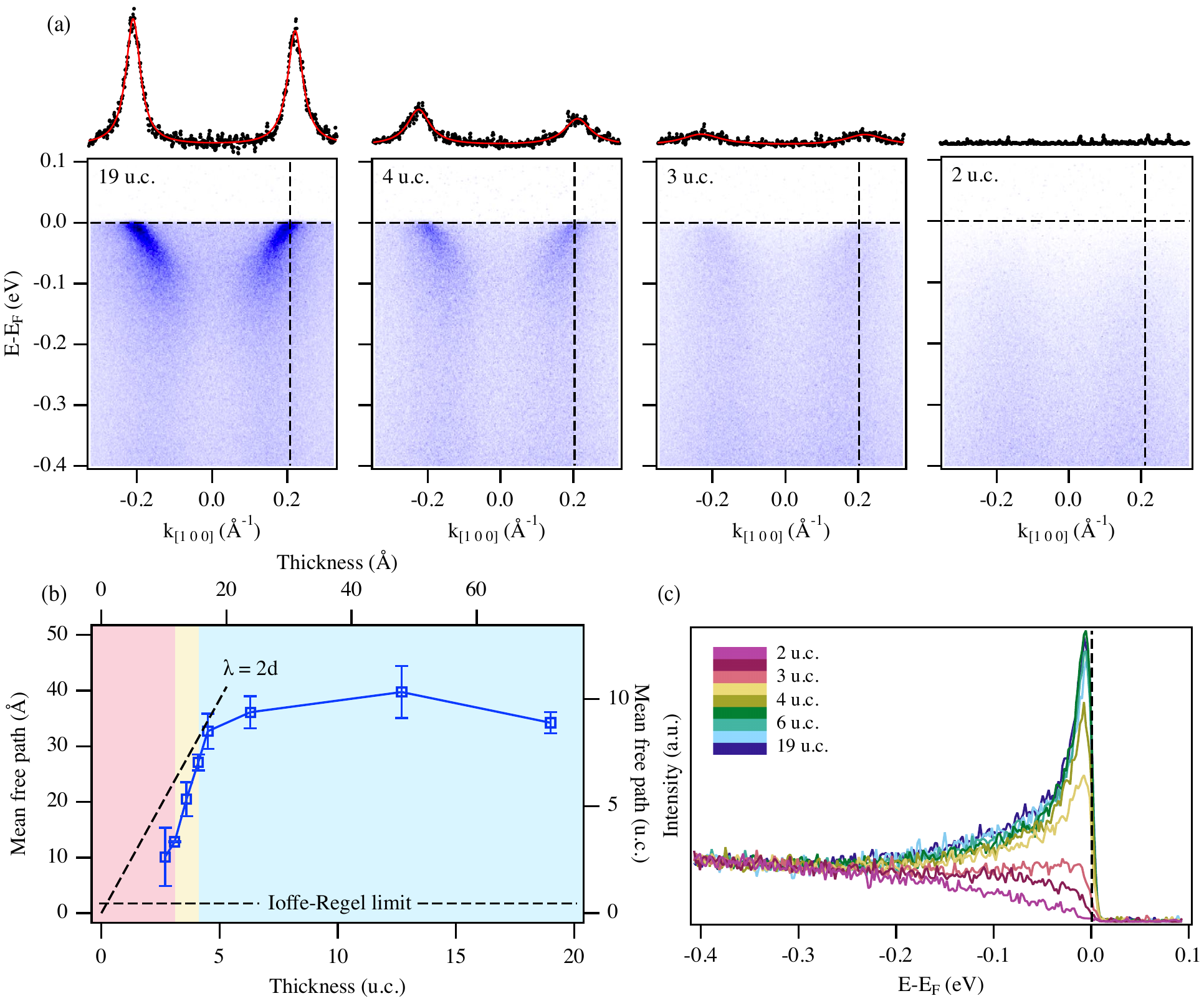}
\caption{\label{fig:arpes}(a) Representative cuts in the electronic band dispersion of LNO thin films of 19, 4, 3 and 2 unit cells. The black profiles are MDCs at the Fermi level. Fits with two Lorentzians are overplotted in red. (b) Carrier mean free path $\lambda=1/\Delta k$ as a function of LNO film thickness. The values have been calculated from the average FWHM of the Lorentzian fits shown in (a) and were corrected for the angle between the $k$-space cut and the Fermi surface contour (for details see main text). The error bars reflect the difference in $\Delta k$ obtained from the two peaks. An estimate of the mean free path corresponding to the Ioffe-Regel limit for LNO is indicated by a horizontal dashed line in the bottom part of the graph. Another dashed line displays a mean free path equal to twice the film thickness, for comparison. The red, yellow, and blue shaded areas correspond to the 3 transport regimes we discuss in the main text. (c) Energy-distribution curves (EDCs) at the Fermi wave vector for all samples in the series.}
\end{figure*}

The photon energy of 11~eV used in our experiments probes the electronic structure along a spherical contour, as depicted in gray in Figure~\ref{fig:fermi_surface}(a). As long as the electronic structure remains 3D, our measurements thus capture the large hole pockets, while no signal is expected from the electron pocket at the zone center, in agreement with the data in \ref{fig:fermi_surface}(b).
This would no longer be the case if the electronic structure were to become 2D and therefore lose all $k_z$ dependence, as reported in Ref.~\cite{yoo2016}. In this case, our measurements should pick up a signal from the electron pocket at $\Gamma$, which is predicted to remain occupied down to a thickness of 2 u.c.~\cite{fowlie2017} The largely unchanged Fermi-surface contour for a film with a thickness of 3~u.c., as shown in Figure \ref{fig:fermi_surface}(c), and the absence of the zone-center electron pocket in this measurement indicate that a transition to 2D behavior does not occur even in ultrathin films.
This discrepancy might be related to the improved $k_z$ resolution of the present study or to factors related to the films themselves, such as surface and interface roughness and film cleanliness, which we expect to influence the effective dimensionality of the electronic structure. We clarify that the electronic dimensionality is different from, albeit related to, the reduced geometric dimensionality dictated by the thinness of the film.



Figure \ref{fig:arpes}(a) shows the electronic band dispersion measured along the dashed line in Figure \ref{fig:fermi_surface}(b). In these measurements, corresponding to films of thickness 19, 4, 3 and 2 u.c., we observe the evolution of the bands which form the hole-like pockets centred at the Brillouin-zone corners. The suppression of intensity near $k_{[100]}=0$ can be attributed to matrix elements and is consistent with an approximately parabolic band closing at $k_{[100]}=0$, as expected theoretically and reported in previous experiments~\cite{king2014}. By examining these cuts, it is evident that the spectral weight at the Fermi edge is progressively lost as the film thickness is reduced and the material turns insulating. On the other hand, the overall band dispersion and the Fermi wave vector $k_F$ remain largely unchanged down to a thickness of 3 u.c., which shows weakly insulating behavior at all temperatures in transport.

Next, we evaluate the mean free path $\lambda$ of the charge carriers as a function of film thickness. The mean free path measured by ARPES is defined as $\lambda=1/\Delta k$, where $\Delta k$ is the full width at half maximum (FWHM) of the peak in a momentum distribution curve (MDC) taken normal to the Fermi surface. To determine $\lambda$ we thus fit the MDCs displayed in Figure \ref{fig:arpes}(a) by Lorentzians (shown in red) and correct their FWHM by a geometric factor accounting for the angle of $\approx45^{\circ}$ between the measured $k$-space cut and the Fermi surface contour.

The values of $\lambda$ extracted by following this procedure are presented in Figure \ref{fig:arpes}(b). The red, yellow, and blue shaded areas correspond to the 3 transport regimes discussed previously. We find that the mean free path follows two distinct regimes. For films up to 5~u.c. thickness, $\lambda$ increases almost linearly from $\approx 10$~\AA{} to $\approx 30$~\AA. In thicker films, $\lambda$ rapidly saturates at values below $\approx 40$~\AA.
We first note that these values exceed the thickness-independent value of 8~\AA{} reported in earlier ARPES experiments~\cite{king2014}. We attribute this predominantly to the improved experimental conditions resulting from a higher spectrometer resolution, the smaller light spot size, which limits the inhomogeneous broadening arising from the average over a large surface area, and the lower photon energy, which reduces the broadening induced by the intrinsically poor $k_z$ resolution of ARPES~\cite{}.
On the other hand, the values of $\lambda$ obtained here remain below the 100~\AA{} deduced from magneto-transport experiments~\cite{scherwitzl2011}. We can presently not exclude that this is due to limitations in the effective resolution of our ARPES experiments which are notoriously hard to quantify. However, this might also be of fundamental origin. The mean free path measured by ARPES contains all scattering events out of a particular $k$-state, while transport is predominantly sensitive to large-angle scattering. ARPES thus generally measures a lower limit of the transport mean free path.
Secondly, we note that the evolution of $\lambda$ naturally explains the puzzling absence of quantum-well states reported in all previous ARPES experiments on ultrathin LNO films, even though they appear in band structure calculations where they show a subband splitting that should be readily resolved in experiment~\cite{king2014, yoo2016, fowlie2017}. Quantum-well states --- and thus a genuinely 2D electronic structure --- can only exist if the coherence length of the electronic states significantly exceeds the film thickness.
As shown in Figure \ref{fig:arpes}(b), this is not the case in our films.
The sharp reduction for the thinnest films limits the coherence length $\lambda$ to below twice the film thickness (indicated by a dashed line in Figure \ref{fig:arpes}(b)), which prohibits the coherent superposition of wave packets reflected at the surface and interface and thus the formation of quantum-well states.
We finally note that the transition from a nearly constant to a sharply decreasing mean free path at a thickness of 5~u.c. coincides with the onset of an upturn in resistivity. This points to a complex interplay of interactions and disorder, rather than to weak localization as the main origin of the resistivity upturn~\cite{alloul2009}. Our ARPES data also show that the mean free path remains well above the Ioffe-Regel limit of $k_F \lambda = 1$ even for the 3~u.c. film, which is insulating in transport, suggesting that the thickness-dependent MIT is not induced by strong localization~\cite{gunnarsson2003, scherwitzl2011, wang2019}.

We now turn our attention to the origin of the sharp reduction in mean free path as the thickness is reduced below 5~u.c. 
At first sight, the nearly linear dependence of the mean free path on film thickness appears to suggest a dominant contribution of elastic defect scattering at the interface and surface. However, while interface roughness and the presence of defects such as oxygen vacancies will inevitably contribute to limiting the mean free path, it cannot readily explain the abrupt change in the thickness dependence of $\lambda$ at 5~u.c. The latter rather points to an important role of fluctuations near the MIT. It has previously been suggested that such fluctuations might be enhanced by the reduced dimensionality of ultrathin films~\cite{boris2011, king2014}. A reduced dimensionality is, however, not confirmed by our data, which show that films down to 3~u.c. do not become electronically 2D.
In contrast to Ref~\cite{king2014}, we believe that fluctuations might also be enhanced by the inevitable structural relaxation of 3D perovskites at interfaces and surfaces. It is known that such relaxations are particularly strong in nickelates and move the outermost unit cells of LNO closer to an insulating state~\cite{kumah2014}, which shares similarities with the ground state of other rare-earth nickelates. In the vicinity of such an insulating state with putative antiferromagnetic order, bond and charge disproportionation, as well as fluctuations of spin, lattice, and charge degrees of freedom should be strong and can have a decisive effect on quasiparticle coherence.
%
This interpretation is further supported by the overall similarity of the spectral function in ultrathin LNO films with earlier observations of NNO films below the MIT~\cite{dhaka2015}. It is, however, at odds with the absence of bond disproportionation and metallic state reported by DFT calculations for a 2-u.c.-thick film in Ref~\cite{fowlie2017}. Most imporantly, both systems show a gradual suppression of spectral weight near the Fermi level, rather than a clear gap
both in optical and photoemission measurements~\cite{dhaka2015, ruppen2015}. 
We thus speculate that the MIT in ultrathin LNO is neither driven primarily by the reduced dimensionality, nor by defects, but rather by structural relaxation of the ultra-thin films. 
 
 
In conclusion, we demonstrated that the combination of sputter deposition and laser ARPES provides high-quality data showing significantly narrower quasiparticle line widths than obtained previously. This establishes sputter deposition as an attractive alternative to MBE or PLD growth for \textit{in-situ} electronic structure studies. Our data revealed a strong dependence of the quasiparticle mean free path on sample thickness, which was  previously unresolved and can explain the universally reported absence of quantum-well states in ARPES experiments on nickelate thin films.
The similarity between the electronic structures of 2-3-u.c.-thick LNO films and insulating NNO further points to an important role of structural relaxation at the interface or surface in the MIT of ultrathin LNO films.

\begin{acknowledgments}
This work has been supported by the Swiss National Science Foundation (Ambizione Grant No. PZ00P2-161327 and project grants 165791, 184998) and by Comunidad de Madrid (Atraccion de Talento grant No. 2018-T1/IND-10521).
\end{acknowledgments}

%

\end{document}